\documentclass[%
superscriptaddress,
 amsmath,amssymb,
 aps,
prl,reprint,nofootinbib, showpacs,showkeys
]{revtex4-1}
\pdfoutput=1
\usepackage{graphicx}
\usepackage{dcolumn}
\usepackage{bm}
\usepackage{slashed}
\usepackage{graphicx}
\usepackage[colorlinks=true,linktocpage=true,linkcolor=blue,citecolor=blue]{hyperref}
\usepackage{float}
\usepackage{nicefrac}
\usepackage[normalem]{ulem}
\usepackage{amsmath}
\usepackage{amstext}
\usepackage{subfigure} 
\usepackage{bbold} 
\usepackage[makeroom]{cancel}
\usepackage{stmaryrd} 
\newcolumntype{R}{>{$}r<{$}}
\newcolumntype{C}{>{$}c<{$}}
\usepackage{colortbl}
\definecolor{jocol}{rgb}{0.1,0.2,0.8}

\definecolor{tgcol}{rgb}{0.7,0.1,0.2}

\newcommand{\beq}{\begin{eqnarray}}
\newcommand{\eeq}{\end{eqnarray}}

\def\LR{\left(} 
\def\RR{\right)}
\def\LS{\left[} 
\def\RS{\right]}

\def\ev{{\boldsymbol e}}

\usepackage{lineno}
\begin{document}
\title{Statistical hadronization model for heavy-ion collisions in the few-GeV energy regime}  
%

	\author{Szymon Harabasz}
	\email{S.Harabasz@gsi.de}
	 \affiliation{Technische Universit\"at Darmstadt, 64289 Darmstadt, Germany}
	
	\author{Wojciech Florkowski}
	\email{wojciech.florkowski@uj.edu.pl}
	 \affiliation{Jagiellonian University,  PL-30-348 Krak\'ow, Poland}
	 
	\author{Tetyana Galatyuk}
	\email{T.Galatyuk@gsi.de}
	 \affiliation{GSI Helmholtzzentrum f\"ur Schwerionenforschung GmbH, 64291 Darmstadt, Germany}
	 	 \affiliation{Technische Universit\"at Darmstadt, 64289 Darmstadt, Germany}
	 
	 \author{Ma\l gorzata Gumberidze}
	\email{M.Gumberidze@gsi.de}
	 \affiliation{GSI Helmholtzzentrum f\"ur Schwerionenforschung GmbH, 64291 Darmstadt, Germany}
	
    \author{Radoslaw Ryblewski} 
    \email{radoslaw.ryblewski@ifj.edu.pl} 
    \affiliation{Institute of Nuclear Physics Polish Academy of Sciences, PL-31342 Krakow, Poland}
    
	\author{Piotr Salabura}
	\email{piotr.salabura@uj.edu.pl}
	 \affiliation{Jagiellonian University,  PL-30-348 Krak\'ow, Poland}
	 
	 \author{Joachim Stroth}
	\email{J.Stroth@gsi.de}
	 \affiliation{GSI Helmholtzzentrum f\"ur Schwerionenforschung GmbH, 64291 Darmstadt, Germany}
	  \affiliation{Institut f\"ur Kernphysik, Goethe-Universit\"at, 60438 Frankfurt, Germany}
\date{\today} 
\bigskip
\begin{abstract}
We show that the transverse-mass and rapidity spectra of protons and pions produced in Au-Au collisions at $\sqrt{s_{\rm NN}}~=~2.4$~GeV can be well reproduced in a thermodynamic model assuming single freeze-out of particles from a spherically symmetric hypersurface. This scenario corresponds to a physical picture used by Siemens and Rasmussen in the original formulation of the blast-wave model. Our framework modifies and extends this approach by incorporation of a Hubble-like expansion of QCD matter and inclusion of resonance decays.  In particular, the $\Delta(1232)$ resonance is taken into account, with a width obtained from the virial expansion. 
Altogether, our results bring evidence for substantial thermalization of the matter produced in heavy-ion collisions in a few GeV energy regime and its nearly spherical expansion.
\end{abstract}
     
\date{\today}  
\maketitle 
\section{Introduction}
Thermal models of hadron production, based on the idea of statistical hadronization, have been very successful in  describing hadron yields in various collision processes, in particular, in heavy-ion collisions (HIC) in a wide range of beam energies and for different projectile-target systems~(see, {\it e.g.}, Refs.~\cite{Cleymans:1992zc,Becattini:1997ii,Florkowski:2001fp,Petran:2013lja,Becattini:2014hla,Vovchenko:2015idt,Andronic:2017pug}). The reasons for studying the thermal aspects of hadron production in heavy-ion collisions are manifold. The hadron abundances can be explained over several orders of multiplicity by fixing just a few thermodynamic parameters. Moreover, the assumption of local thermalization of the expanding dense and hot matter formed in the collision (called a fireball) allows to apply hydrodynamic concepts  \cite{Florkowski:2017olj,Romatschke:2017ejr} for describing its evolution and the emission of electromagnetic radiation~\cite{Adamczewski-Musch:2019byl}. Such an approach has been very successful in the description of HIC at ultra-relativistic energies and helped to identify landmarks in the QCD phase diagram in the region of vanishing net-baryon density, which is also accessible by lattice QCD calculations~\cite{Borsanyi:2010bp,Bazavov:2018mes}. 

HIC at lower beam energies provide access to strongly interacting matter at high net-baryon densities where a rich structure in the QCD phase diagram is expected but lattice QCD is not applicable. The problem if the fireball formed in a few GeV beam energy range (where in central collisions essentially all nucleons are stopped in the center-of-mass frame) is thermalized remains still a matter of debate~\cite{Lang:1990zc,Endres:2015fna,Galatyuk:2015pkq,Staudenmaier:2017vtq}.
The study of hadron spectra is crucial to answer this question. In a thermal analysis, however, it has to be first demonstrated that the experimental hadron yields can be well described with a few thermodynamic parameters such as temperature, $T$, and the baryon chemical potential, $\mu_B$. Only in the second step, the transverse-mass spectra, which are typically falling off exponentially, have to be reproduced. One should note, however, that collective radial expansion (specified  by the flow~$v$) and resonance decays also affect the momentum distribution of hadrons~\cite{Schnedermann:1993ws}. 

The two physical aspects mentioned above are unified in a single-freeze-out model~\cite{Broniowski:2001we,Broniowski:2002wp}, which identifies the chemical and kinetic freeze-outs by neglecting hadron re-scattering processes (after the chemical freeze-out). This model assumes a sudden freeze-out governed by local thermodynamic conditions. This concept is implemented in the THERMINATOR Monte-Carlo generator~\cite{Kisiel:2005hn,Chojnacki:2011hb}, which allows for studies of hadron production taking place on arbitrary freeze-out hypersurfaces defined in the four-dimensional space-time. The most popular parametrization of such a freeze-out hypersurface~\cite{Schnedermann:1993ws}, dubbed the blast-wave model, assumes the symmetry of boost invariance (along the beam axis). As a matter of fact, it was introduced as a modification of the original blast-wave model formulated by Siemens and Rasmussen (SR)~\cite{Siemens:1978pb}, which instead of the  boost invariance employed a spherical symmetry of the freeze-out geometry. 

We present herein a novel approach towards consistent simultaneous description of hadron yields and transverse-mass spectra with the SR model. This approach offers an alternative interpretation of experimental results based on a concept of thermal equilibrium, as compared to commonly used transport model approaches. 
The spherical symmetry of a fireball may be natural at low energies, where the colliding nuclei are definitely not transparent to each other (the energy dependence of this effect is shown  in Ref.~\cite{Bearden:2003hx}). In any case, compared to the boost invariance, the spherical symmetry seems to be a better starting point for the description of HIC in a few GeV energy regime and we are going to verify this concept in this work.  In order to analyze data collected for Au-Au collisions at $\sqrt{s_{\rm NN}}= 2.4$~GeV by the HADES collaboration, we implement the SR model into the THERMINATOR Monte-Carlo framework. This allows for a more comprehensive study compared to those done previously~\cite{Schuldes:2016eqz}. To select a reaction class where thermalization is most likely to occur, we focus on central collisions only. 

\section{Siemens-Rasmussen model} 
The basis for this model is the Cooper-Frye formula~\cite{Cooper:1974mv} that describes the invariant momentum spectrum of particles emitted from an expanding source
\begin{equation}
E_p \frac{dN}{d^3p} = \int d^3\Sigma(x) \cdot p \,\, f(x,p).
\label{eq:CF1}
\end{equation}
Here $f$ is the phase-space distribution function of particles, $p$ is their four-momentum with the mass-shell energy,  $p^0 = E_p = \sqrt{m^2 + \boldsymbol{p}^2}$, and $d^3\Sigma_\mu(x)$ is the element of a three-dimensional freeze-out hypersurface from which particles are emitted.\footnote{Three-vectors are shown in bold font and a dot is used to denote the scalar product of both four- and three-vectors, $g_{\mu\nu} = \hbox{diag}(+1,-1,-1,-1)$.} 

Herein we adopt the simplest form of a spherically symmetric freeze-out defined by the space-time coordinates
\begin{equation}
x^\mu = (t,x,y,z) = \LR t(r), r\, \ev_r \RR,
\label{xmurad} 
\end{equation}
where  $\ev_r = \LR  \cos\phi\sin\theta,  \sin\phi \sin\theta,  \cos\theta \RR$. Here $\phi$ and $\theta$ are the azimuthal and polar angles relative to the beam axis, respectively, and $t(r)$ defines the freeze-out times, {\it i.e.}, the times when the hadrons in the shells of radius $r$ stop to interact ($0 \leq r \leq R$). Below, we assume sudden freeze-out of the expanding fireball with $t(r)=$ const, which implies
\begin{eqnarray}
\! d^3\Sigma_\mu \!&=& \!
\LR 1, 0,0,0  \RR
r^2  \sin\theta \,d\theta
\,d\phi\,dr.
\label{d3sigmarad}
\end{eqnarray}
Besides the spherically symmetric hypersurface,  we introduce a spherically symmetric (hydrodynamic) flow
\begin{eqnarray}
u^\mu &=& \gamma(r)\LR 1,
v(r)\ev_r  \RR,
\label{umurad1}
\end{eqnarray}
where $\gamma(r) = (1-v^2(r))^{-1/2}$ is the Lorentz factor. 
With the hadron four-momentum defined as $p^\mu = \LR E_p, p\, \ev_p \RR$, where $\ev_p = \LR  \cos\varphi \sin\vartheta, \sin\varphi \sin\vartheta,  \cos\vartheta \RR$, one can easily find that $p \cdot u=\gamma \LR E_p\!-\!p v \kappa \RR$, where $\kappa = \cos\theta \cos\vartheta + \sin\theta \sin\vartheta \cos(\phi-\varphi) $, and
\begin{equation}
d^3\Sigma \cdot p = E_p  r^2 \sin\theta
\,d\theta \, d\phi \, dr.
\label{Sigmaprad}
\end{equation}

\section{Local equilibrium.}  
In this work we assume that the hadron system formed at freeze-out is very close to local thermodynamic equilibrium, hence, the distribution function $f(x,p)$ has the form
\begin{equation}
f(x,p) = \frac{g_s}{(2\pi)^3}\LS \Upsilon^{-1} \exp\LR\frac{p \cdot u}{T}\RR-\epsilon \RS^{-1},
\label{eq:eq}
\end{equation}
where  $\epsilon=-1$ ($\epsilon=+1$) for Fermi-Dirac (Bose-Einstein) statistics and $g_s = 2s +1$ is the spin degeneracy factor. 
Please note that local thermalization at freeze-out does not exclude the existence of substantial pressure anisotropies of the system at earlier stages, as suggested, for example, in Ref.~\cite{Lang:1990zc}.
The fugacity $\Upsilon$ is defined as~\cite{Torrieri:2004zz}
\begin{eqnarray}
\Upsilon=  
 \gamma_q^{N_q+N_{\bar q}} \gamma_s^{N_s+N_{\bar s}}  \exp \left( \frac{\mu}{T}\right),
\label{upsiNeq}
\end{eqnarray}
where $\mu=\sum_{Q} Q \mu_Q$, with $Q$ denoting the conserved quantum numbers for each hadron, $Q\in \{B,I_3,S\}$. The parameters $\gamma_q$ and $\gamma_s$ are included to account for deviations from chemical equilibrium, while $N_q$ and $N_s$ ($N_{\bar q}$ and $N_{\bar s}$) denote the numbers of light and strange quarks (antiquarks) in the hadron. In the case of grand canonical ensemble with chemical equilibrium, one sets $\gamma_q=\gamma_s=1$. To allow for strangeness under-saturation, a characteristic feature of the particle spectra observed at beam energies discussed here, we allow $\gamma_s$ to be smaller than unity but keep $\gamma_q = 1$.

We stress that our framework includes in Eq.~(\ref{eq:CF1}) all the contributions from decays of heavier resonances, although, in contrast to high-energy collisions studied at RHIC and the LHC, most of them are very small or negligible. The dominant contribution,   in addition to the pions born on the freeze-out hypersurface and called ``primordial'', comes from decays of the lowest-lying baryonic resonance, {\it i.e.}, $\Delta(1232)$.  For a proper treatment of the decay pions, the inclusion of the $\Delta(1232)$ width is important. This is achieved by using the  density function obtained in Ref.~\cite{Lo:2017ldt} from the pion-nucleon phase shift in the $P_{33}$ channel (see also \cite{Weinhold:1997ig}). 
\begin{figure*}[t]
\begin{center}
\includegraphics[width=\textwidth]{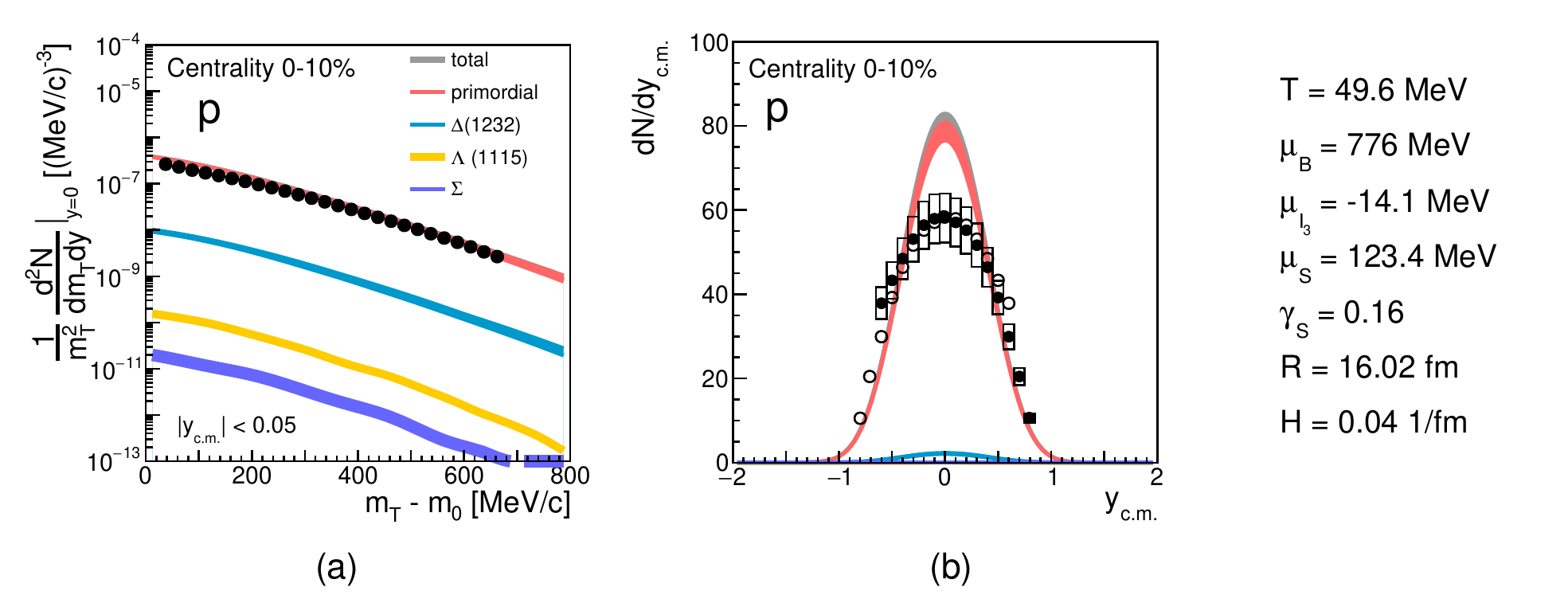}
\end{center}
\caption{Distributions of protons: rapidity (a) and transverse-mass ($m_T$) at mid-rapidity (b). 
The black dots describe the experimental values used in the fit of $H$. 
Bands describe model predictions for the optimal primordial and secondary contributions. 
The band widths reflect the uncertainty of theoretical predictions connected with the experimental errors.
The model results are corrected for bound protons such that the spectrum reproduces the measured proton multiplicity.
Experimental data points are from \cite{Szala:2019}, in panel (a) full circles represent measured data, open circles -- reflected around mid-rapidity, rectangles -- systematic uncertainties.}
\label{fig:proton}
\end{figure*}
%
\section{Hubble-like radial flow}
In the original SR blast-wave model~\cite{Siemens:1978pb}, it is assumed that the thermodynamic parameters as well as the radial flow velocity are constant ($T = \hbox{const},\mu = \hbox{const}, v = v_0 = \hbox{const}$). 
The condition of constant radial flow breaks the natural requirement that the flow at the center of the system should vanish, $v(r=0)=0$. 
Moreover, results of full hydrodynamic calculations indicate that the radial flow linearly grows with $r$ for small values of $r$ and approaches unity ({\it i.e.}, the speed of light) in the limit $r \to \infty$~\cite{Chojnacki:2004ec}. 
These observations suggest that one can use the flow parametrization $v(r) = \tanh(H r)$, where $H$ is a constant. 
For small values of $r$ we have $v \sim H r$, hence $r$ plays a role of the Hubble constant. 

\section{Fitting strategy and comparison with the HADES data.} 
\begin{table}[b]
\begin{center}
\begin{tabular}{ CCCC }
\hline \hline
    \text{particle}  & \text{multiplicity} \:\:\: & \text{uncertainty} & \text{Ref.} \\ \hline 
    p      & 77.6   & \pm 2.4 & \text{\cite{Szala:2019,Szala2019a}}\\
    p \: \text{(bound)} & 46.5 &  \pm  1.5 &\text{\cite{Szala:2019,Szala2019a}}\\
    \pi^+  & 9.3   &  \pm 0.6 & \text{\cite{Gumberidze:2019}}\\
    \pi^-  & 17.1   & \pm 1.1 & \text{\cite{Gumberidze:2019}}\\
    K^+    & 5.98\,10^{-2} & \pm 6.79\,10^{-3}  & \text{\cite{Adamczewski-Musch:2017rtf}} \\
    K^-    & 5.6\,10^{-4} & \pm 5.96 \,10^{-5} & \text{\cite{Adamczewski-Musch:2017rtf}} \\
    \Lambda & 8.22\,10^{-2} & ^{+5.2} _{-9.2}\,10^{-3} & \text{\cite{Adamczewski-Musch:2018xwg}} \\ 
    \hline \hline
\end{tabular}

\caption{Particle multiplicities used in the determination of the freeze-out parameters. Protons bound in nuclei are taken into account as shown. \label{tab:mult}}
\end{center}
\end{table}

In the first step we obtain thermodynamic model parameters from the ratios of experimental multiplicities measured by HADES in the full phase space for the 10\% most central Au-Au collisions~\cite{Szala:2019,Szala2019a,Gumberidze:2019,Adamczewski-Musch:2017rtf,Adamczewski-Musch:2018xwg}. The analyzed ratios include protons, positively and negatively charged pions, positively and negatively charged kaons, and $\Lambda$-hyperons, as listed in Tab.~\ref{tab:mult}. In this calculation, we assume that the protons finally bound in the emitted deuterons, tritons, and $^3$He nuclei~\cite{Szala:2019,Szala2019a} initially freeze out as unbound nucleons, hence they are included in the proton yield (see~Tab.\ \ref{tab:mult}). Our analysis gives the following values of the thermodynamic parameters: $T$=~49.6~$\pm$~1~MeV, $\mu_B$=~776~$\pm$~3~MeV, $\mu_{I_3}=-14.1 \pm 0.2$~MeV, $\mu_S$=~123.4~$\pm$~2~MeV, and $\gamma_s$=~0.16$\pm$~0.02, where the errors were estimated from calculations with the multiplicities varied within given errors (see again Tab. \ref{tab:mult}).   
These results are close to those found and discussed in Refs.~\cite{Cleymans:2005xv,Castorina:2019vex}.


For a fixed value of $H$, the absolute normalization of the yields determines the value of $R$. Hence we may treat $R$ as a function of $H$ and we are left with only one independent parameter, {\it i.e.}, $H$. Its value is obtained from the fit of the proton transverse-mass spectrum  by minimization of the quadratic deviation
\begin{equation}
Q^2(H) = \sum_i \frac{\left(Q_{i, \rm model}(H)-Q_{i, \exp}\right)^2}{Q^2_{i, \exp}},
\label{eq:optH}
\end{equation}
where $Q_{i, \rm model}(H)$ and $Q_{i, \exp}$ denote the model and experimental values. 
In Eq.~(\ref{eq:optH}) all the points from the experimentally available proton spectrum are included. 
The minimization procedure yields the value $H =$~0.04~fm$^{-1}$ (with the corresponding radius $R =$~16.02~fm and the mean radial flow $\langle v \rangle \sim 0.4$). 
Using this result, we obtain a very good agreement ($Q = 0.20$) between the data and model predictions as shown in Fig.~\ref{fig:proton}~(a).
\begin{figure*}[t]
\begin{center}
\includegraphics[width=\textwidth]{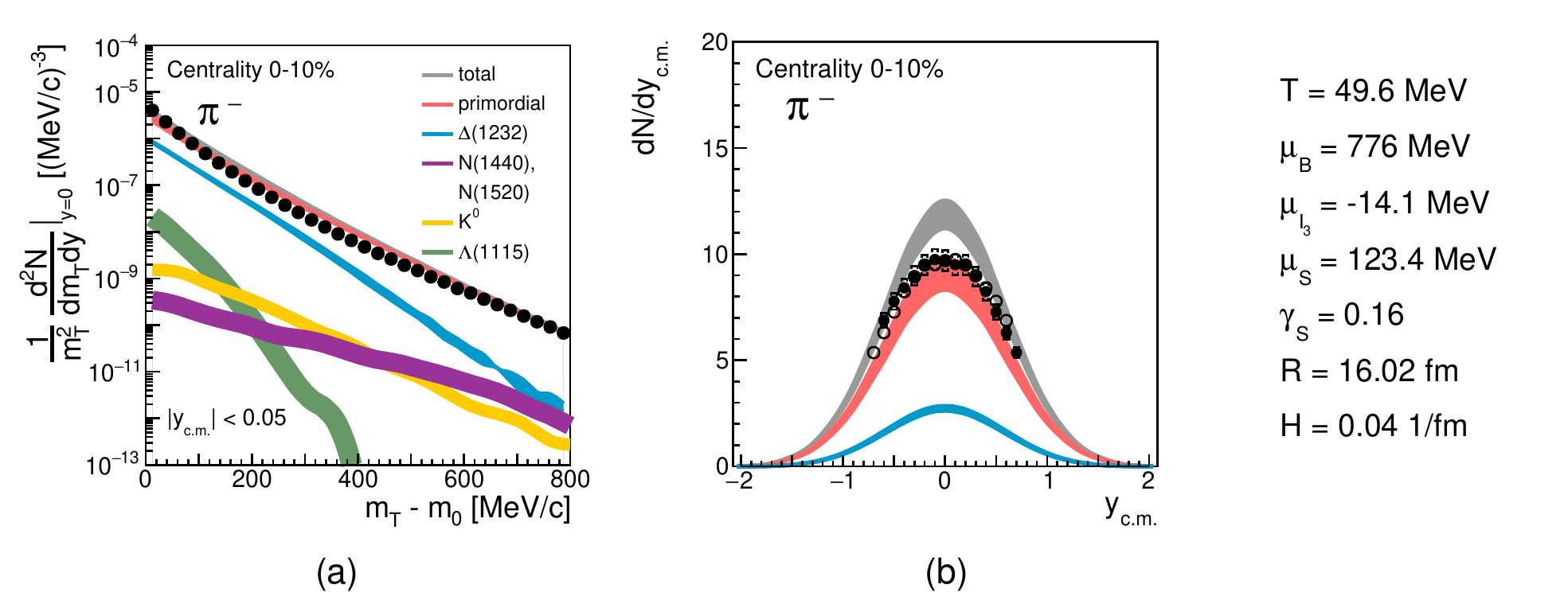}
\end{center}
\caption{Same as Fig.~\ref{fig:proton} but for negatively charged pions. }
\label{fig:pim}
\end{figure*}
Having determined the value of $H$, we can calculate other model spectra. 
In particular, we can compare the proton rapidity distribution obtained from the model and check if it consistently well describes the data (along with the transverse-mass spectrum). 
Our results are shown in Fig.~\ref{fig:proton}~(b). 
They indicate that the model distribution is too narrow, with a theoretical value exceeding the data by about 30\% at $y=0$.
Nevertheless, the data points for $y > 0.4$ agree well with the model curve and we find $Q = 0.28$ for the proton rapidity spectrum alone.\footnote{We stress that except for the transverse-mass spectrum of protons, other values of $Q$ are predictions of the model calculation.} The fact that the rapidity distribution is equally well described (compared to the transverse-mass distribution) points out the approximate spherical symmetry of the produced system. 

Our approach reproduces the main features of the proton data at the quantitative level, and confirms our original conjecture that spherical symmetry is a good assumption for the description of systems produced in central collisions of HIC at low energies. Clearly, the boost-invariant blast-wave models (yielding constant $dN/dy$) are not capable of reproducing the Gaussian shape of the experimental rapidity distribution as that shown in Fig.~\ref{fig:proton}~(b), see also~\cite{Schuldes:2016eqz}.

%
%

Figure \ref{fig:pim} shows our results for negatively charged pions. 
It is important to emphasize that these results are obtained with the parameters fixed by the hadron ratios and the proton transverse-mass spectrum, hence, there is no room for extra model adjustments in these cases.
For the negatively charged pions (as well as for the positively charged ones that are not shown here) we observe a good agreement between the model and experimental spectra. We obtain $Q=0.46$ ($Q=0.28$) for the transverse-mass distributions for  negatively (positively) charged pions, and $Q=0.12$ ($Q=0.16$) for the corresponding rapidity distribution, respectively. 
Interestingly, the rapidity distributions are better reproduced compared to the transverse-mass distributions.
The quality of agreement between transport models and the HADES data on pion production has been recently reported in \cite{Gumberidze:2019}. Our simple model provides comparable (or even better) description of experimental data.
Similar quality of description is obtained for rapidity distributions of particles containing strangeness but somewhat worse for their transverse-mass spectra. This will be discussed in a forthcoming paper.  We note that the model rapidity distributions for all hadrons are too narrow, which indicates that the systems created in HIC at this energy are more elongated along the beam axis, as compared to the model assumptions. This behavior suggests an incomplete stopping\footnote{It is important to note that the range of impact parameters included in the $0-10\%$ centrality class reaches out to about 3~fm.}.

As already mentioned, our framework includes feed-down contributions to the hadron spectra from (strong) resonance decay. We find that the most important effect comes from the $\Delta(1232)$ decay which significantly contributes to the pion spectra. The relative contribution can  easily be assessed by comparing the red and grey bands in Fig.~\ref{fig:pim}.  The contributions from other resonances are negligible, at least within the limits of precision of our model description.

\section{Summary and Conclusions} 
In this work we have studied the rapidity and transverse-mass spectra of protons and pions produced in Au-Au collisions at $\sqrt{s_{\rm NN}} = 2.4$~GeV. 
We have found that they can be well reproduced in a thermodynamic model that assumes single freeze-out of hadrons from a spherically symmetric hypersurface.
Such a spherical geometry was used by Siemens and Rasmussen in their original formulation of the blast-wave model.
Our framework modifies and extends this approach by incorporation of the Hubble-like expansion of matter and inclusion of the resonance decays.
We have found that the presence of the $\Delta$ resonance affects the spectra of pions, while the contributions from other resonances can be neglected.
The obtained thermodynamic parameters agree well with the universal freeze-out curve.
Altogether, our results bring  evidence for substantial thermalization of the matter produced in a few GeV energy range and its nearly spherical expansion.
These are remarkable findings and they pave a way for future supplements to our approach.

\section{Acknowledgements} This work was supported in part by: the Polish National Science Center Grants No.~2016/23/B/ST2/00717, No.~2018/30/E/ST2/00432, and No.~2017/26/M/ST2/00600; TU Darmstadt, Darmstadt (Germany), Helmholtz VH-NG-823, DAAD PPP Polen 2018/57393092; Goethe-University, Frankfurt(Germany), and HIC for FAIR (LOEWE).

%

\end{document}